# The Heisenberg versus the Schrödinger picture and the problem of gauge invariance.

by

Dan Solomon

Rauland-Borg Corporation

3450 W. Oakton Street

Skokie, IL 60076

Phone: 847-324-8337

Email: **dan.solomon@rauland.com**

**Abstract**

It is generally assumed that quantum field theory (QFT) is gauge invariant. However it is well known that non-gauge invariant terms appear in various calculations. This problem was recently examined in [9] for a "simple" field theory and it was shown that for this case QFT in the Schrödinger picture is not, in fact, gauge invariant. In order to shed further light on this problem we will examine the Heisenberg and Schrödinger formulations of QFT. It is generally assumed that these two "pictures" are equivalent; however we will show that this is not necessarily the case. We shall consider a "simple" field theory consisting of a quantized fermion field in the presence of a classical electromagnetic field. We will show that, although the two pictures are formally equivalent, the Heisenberg picture is gauge invariant but that the Schrödinger picture is not. This suggests that the proper way to formulate QFT is to use the Heisenberg picture.

## 1. Introduction

Quantum field theory is assumed to be gauge invariant [1][2]. A change in the gauge is a change in the electromagnetic potential that does not produce a change in the electromagnetic field. The electromagnetic field is given by,

$$\vec{E} = -\left(\frac{\partial \vec{A}}{\partial t} + \vec{\nabla} A_0\right); \quad \vec{B} = \vec{\nabla} \times \vec{A} \tag{1.1}$$

where $\vec{E}$ is the electric field, $\vec{B}$ is the magnetic field, and $(A_0, \vec{A})$ is the electromagnetic potential. A change in the potential that does not produce a change the electromagnetic field is given by,

$$\vec{A} \to \vec{A}' = \vec{A} - \vec{\nabla}\chi, \quad A_0 \to A_0' = A_0 + \frac{\partial \chi}{\partial t} \tag{1.2}$$

where $\chi(\vec{x}, t)$ is an arbitrary real valued function.

In order for quantum field theory to be gauge invariant a change in the gauge cannot produce a change in any physical observable such as the current and charge expectation values. However, it is well known that when certain quantities are calculated using standard perturbation theory the results are not gauge invariant. For example, the first order change in the vacuum current, due to an applied electromagnetic field, can be shown to be given by,

$$J^\mu_{vac}(x) = \int \pi^{\mu\nu}(x - x') A_\nu(x') d^4 x' \tag{1.3}$$

where $\pi^{\mu\nu}$ is called the polarization tensor and summation over repeated indices is assumed. The above relationship is normally written in terms of Fourier transformed quantities as,

$$J^\mu_{vac}(k) = \pi^{\mu\nu}(k) A_\nu(k) \tag{1.4}$$

where k is the 4-momentum of the electromagnetic field. In this case, using relativistic notation, a gauge transformation takes the following form,

$$A_\nu(k) \to A'_\nu(k) = A_\nu(k) + i k_\nu \chi(k) \tag{1.5}$$

The change in the vacuum current, $\delta_g J^\mu_{vac}(k)$, due to a gauge transformation can be obtained by using (1.5) in (1.4) to yield,

$$\delta_g J^\mu_{vac}(k) = i k_\nu \pi^{\mu\nu}(k) \chi(k) \tag{1.6}$$

Now the vacuum current is an observable quantity therefore, if quantum theory is gauge invariant, the vacuum current must not be affected by a gauge transformation. Therefore $\delta_g J_{vac}^\mu(k)$ must be zero. For this to be true we must have that,

$$k_\nu \pi^{\mu\nu}(k) = 0 \qquad (1.7)$$

However, a review of the literature will easily show that when the polarization tensor is calculated it is found that the above relationship does not hold.

Consider, for example, a calculation of the polarization tensor by W. Heitler (see page 322 of [3]). Heitler's solution for the Fourier transform of the polarization tensor is,

$$\pi^{\mu\nu}(k) = \pi_G^{\mu\nu}(k) + \pi_{NG}^{\mu\nu}(k) \qquad (1.8)$$

The first term on the right hand side is given by,

$$\pi_G^{\mu\nu}(k) = \left(\frac{2e^2}{3\pi}\right)\left(k^\mu k^\nu - g^{\mu\nu}k^2\right)\int_{2m}^\infty dz \frac{\left(z^2 + 2m^2\right)\sqrt{\left(z^2 - 4m^2\right)}}{z^2\left(z^2 - k^2\right)} \qquad (1.9)$$

where m is the mass of the electron, $e$ is the electric charge, and $\hbar = c = 1$. This term is gauge invariant because $k_\nu \pi_G^{\mu\nu} = 0$. The second term on the right of (1.8) is

$$\pi_{NG}^{\mu\nu}(k) = \left(\frac{2e^2}{3\pi}\right) g_\nu^\mu \left(1 - g^{\mu 0}\right)\int_{2m}^\infty dz \frac{\left(z^2 + 2m^2\right)\sqrt{\left(z^2 - 4m^2\right)}}{z^2} \qquad (1.10)$$

where there is no summation over the two $\mu$ superscripts that appear on the right. Note that $\pi_{NG}^{\mu\nu}$ is not gauge invariant because $k_\nu \pi_{NG}^{\mu\nu} \neq 0$. Therefore to get a physically valid result it is necessary to "correct" equation (1.8) by dropping $\pi_{NG}^{\mu\nu}$ from the solution.

Another example of a calculation of the polarization tensor is given by J.J. Sakurai (See pages 273-275 of [4]). Sakurai shows that, based on considerations of Lorentz covariance, the polarization tensor must have the form,

$$\pi_{\mu\nu}(k) = D\delta_{uv} + k^2 \delta_{uv} \pi^{(1)}(k^2) + k_u k_\nu \pi^{(2)}(k^2) \qquad (1.11)$$

where $D$ is a constant. (Note that the use $\delta_{uv}$ instead of $g_{uv}$ reflects the notational conventions of [4]). In order for the above expression to be gauge invariant $D$ must be zero. However Sakurai shows that $D$ is given by the expression,

$$D = ie^2 \int \frac{d^4p}{(2\pi)^4} \frac{(2p^2 + 4m^2)}{(p^2 + m^2 - i\varepsilon)^2} \tag{1.12}$$

Concerning the constant $D$, Sakurai writes "It is not difficult to convince oneself that almost any 'honest' calculation gives $D \ne 0$. In fact ... one can readily shown that $D$ is a positive, real constant..." (see page 275 of [4]). Therefore the result that Sakurai achieves for the polarization tensor is not gauge invariant. In order to make the result gauge invariant the quantity $D$ must be removed.

Another example of a calculation of the polarization tensor is given by K. Nishijima (see section 6-4 of Ref. [5]). He shows that the expression that he obtains for the polarization tensor is not gauge invariant as calculated. In order to obtain a gauge invariant expression the non-gauge invariant part of the expression must be removed. (See discussion after Eq. 6-79 of Ref. [5]).

A similar situation exists when other sources in the literature are examined. For example consider the discussion in Section 14.2 of Greiner et al [2]. Greiner et al write the solution for the polarization tensor (see equation 14.43 of [2]) as,

$$\pi^{\mu\nu}(k) = (g^{\mu\nu}k^2 - k^\mu k^\nu)\pi(k^2) + g^{\mu\nu}\pi_{sp}(k^2) \tag{1.13}$$

where the quantities $\pi(k^2)$ and $\pi_{sp}(k^2)$ are given in [2]. Referring to (1.7) it can be easily shown that the first term on the right is gauge invariant. However the second term is not gauge invariant unless $\pi_{sp}(k^2)$ equals zero. Greiner et al show that this is not the case. Concerning this term Greiner et al write (page 398 of [2]) "… this latter term violates the gauge invariance of the theory. This is a very sever contradiction to the experimentally confirmed gauge independence of QED. [This problem indicates] that perturbative QED is not a complete theory. As one counter example or inconsistency suffices to prove a theory wrong, we should, in principle, spend the rest of this book searching for an improved theory. However, there is little active work on this today because: (1) there is a common belief that some artifact of the exact mathematics is the source of the problem; (2) this problem may disappear when the properly generalized theory, including in its framework all charged Dirac particles, is achieved." In order to achieve a gauge invariant result Greiner et al drop the quantity $\pi_{sp}(k^2)$ from the expression for the polarization tensor.

For another example consider the expression for the polarization tensor as derived by Greiner and Reinhardt (See Section 5.2 of [6]). The polarization tensor is given in Eq. 5.7 of [6] as,

$$\frac{i\pi_{\lambda\sigma}(k)}{4\pi} = -e^2 \int d^4p Tr\left[\gamma_\lambda \frac{1}{(\not{p}-m+i\varepsilon)}\gamma_\sigma \frac{1}{(\not{p}-\not{k}-m+i\varepsilon)}\right] \quad (1.14)$$

They show that this quantity is not gauge invariant. In order to obtain a gauge invariant expression they use Pauli-Villars regularization [7]. They modify the above expression by adding the quantity,

$$-e^2 \int d^4p \sum_{i=1}^{N} C_i Tr\left[\frac{\gamma_\lambda(\not{p}+M_i)\gamma_\sigma(\not{p}-\not{k}+M_i)}{(p^2-M_i^2+i\varepsilon)((p-k)^2-M_i^2+i\varepsilon)}\right] \quad (1.15)$$

According to the Pauli-Villars procedure the auxiliary masses $M_i$ and constants $C_i$ are adjusted so that the non-gauge invariant terms are cancelled. This procedure removes the offending terms, however there is no physical process that justifies this step. That is, the auxiliary masses are not presumed to correspond to actual physical particles. They are simply a mathematical device that is used to get a physically correct result. Therefore, we see that the original calculation is not gauge invariant and must be corrected by the application of an additional step which was not part of the original formulation of the theory.

For another example refer to equation 7.79 of Peskin and Schroeder [8]. Here they show that the expression that they obtain for the polarization tensor is not gauge invariant. In order to obtain a gauge invariant expression they introduce an additional step called dimensional regularization. This "corrects" the problem by removing the unwanted terms from the expression but, as was the case with Pauli-Villars regularization, it is at the expense of introducing a procedure that was not a part of the original formulation of the theory.

Therefore, we see from this review of the literature, that when the polarization tensor is calculated the result is not gauge invariant. The non-gauge invariant part of the result must be removed in order to achieve a physically acceptable result. This removal can be done by "hand" or by the use of an additional mathematical step called regularization. The obvious question to ask, then, is why does this problem occur? If the theory is gauge invariant why does a calculation of the polarization tensor produce non-gauge invariant terms?

This question was examined in some detail in Refs. [9] and [10]. In these papers the problem of gauge invariance was examined for a "simple" field theory in the Schrödinger picture consisting of a quantized fermion field in the presence of an unquantized classical electromagnetic field. In [9] four elements that are normally considered to be part of quantum field theory were examined. These were that (1) the Schrödinger equation governs the dynamics of the theory with the Hamiltonian specified by Eq. (2.2) of [9]; (2) the theory is gauge invariant; (3) there is local charge conservation, i.e., the continuity equation is true; (4) there is lower bound to the free field energy. It was shown that these elements of QFT are not mathematically consistent. Specifically item (2) is incompatible with item (4), that is, if QFT is gauge invariant then there cannot be a lower bound to the free field energy. However it can be readily shown that the vacuum state is a lower bound to the free field energy. Therefore, as discussed in [9], QFT in the Schrödinger picture is not gauge invariant at the formal level. This, then, explains why non-gauge invariant terms appear in the polarization tensor. Since the theory is not gauge invariant in the first place it would be expected that the results of calculation are also not gauge invariant. A similar conclusion was obtained in [10].

The conclusion of this research was that there is a mathematical inconsistency in QFT regarding the way the vacuum state is defined. That is, the vacuum state is defined in a way that is not compatible with the requirements of gauge invariance. It is the purpose of this paper to continue this discussion and see how this inconsistency affects other aspect of the theory. In particular we will examine the relationship between the Schrödinger and Heisenberg pictures. We will examine the "simple" field theory discussed in [9] and [10] in the Schrödinger picture and compare this to the Heisenberg picture. These two "pictures" are generally assumed to be equivalent; however we will show that this is not the case for the field theory under consideration. It will be demonstrated that, even though the two pictures can be shown to be formally equivalent, they yield different results when actual problems are worked out. It will be shown that Heisenberg picture is gauge invariant but that the Schrödinger picture is not. This suggests that if QFT was formulated along the lines of the Heisenberg picture instead of the Schrödinger picture the problems of gauge invariance would be resolved.

The possibility that the Heisenberg and Schrödinger pictures are not equivalent was advocated by P.A.M Dirac in a paper with the interesting title "Quantum Electrodynamics without Dead Wood" [11] (see also [12]). The "dead wood" in this case is the vacuum to vacuum transitions that are part of perturbation theory in the Schrödinger picture. Dirac analyses a "toy model" field theory and creates a situation for which solutions exist in the Heisenberg picture but solutions do not exist in the Schrödinger picture. He uses this result to support his argument that the two pictures are not equivalent and that the Heisenberg picture is the correct approach. In the present paper we will reach a similar conclusion although the approach to the problem taken here is considerably different then that of Dirac.

The paper will proceed as follows. In Section 2 the different approaches leading to the Heisenberg and Schrödinger pictures will be discussed. In Section 3 the Heisenberg picture is developed and easily shown to be gauge invariant. In Section 4 the Schrödinger picture is described. The vacuum state and Schrödinger picture field operator are then defined in Section 5. We define the free field energy as the energy of the system when the electromagnetic potential is zero. It is shown that in the Schrödinger picture the free field energy of any state must be greater than or equal to the free field energy of the vacuum state. In Section 6 we examine gauge invariance in the Schrödinger picture. In contrast to the Heisenberg picture it is shown that the Schrödinger picture is *not* gauge invariant. In section 7 it is shown that in the Heisenberg picture there is no lower bound to the free field energy which is contrast to the Schrödinger picture where there is a lower bound. These results are then summarized in Section 8.

## 2. Heisenberg versus Schrödinger picture.

In quantum field theory a quantum system, at a given point in time, is specified by the state vector $|\Omega\rangle$ and field operator $\hat{\psi}(\vec{x})$. We will write this as the pair $(|\Omega\rangle, \psi)$.

Let the state vector $|\Omega\rangle$ and the field operator $\hat{\psi}(\vec{x})$ be defined at some initial point in time, say $t=0$. This may be taken as the initial conditions of the quantum system. Now there are two ways to handle the time evolution of the system. In the Schrödinger picture it is assumed that field operator $\hat{\psi}(\vec{x})$ is constant in time and the time dependence of the system goes with the state vector $|\Omega(t)\rangle$. In the Heisenberg picture the time dependence is assigned to the field operator $\hat{\psi}(\vec{x},t)$ and the state vector $|\Omega\rangle$ remains constant in time. It is generally assumed that both pictures give equivalent results in that the expectation values of operators are the same. However we will show, in the following discussion, that this is not true.

Note that at the initial time, $t=0$, both pictures are identical. Therefore the time independent Schrödinger field operator $\hat{\psi}(\vec{x})$ is equal to $\hat{\psi}(\vec{x},0)$, which is the time dependent Heisenberg field operator at $t=0$. Similarly, the time independent Heisenberg state vector $|\Omega\rangle$ equals $|\Omega(0)\rangle$, which is the time dependent Schrödinger state vector $|\Omega(t)\rangle$ at $t=0$. For example, let the initial state of the system, at $t=0$, be represented by the pair $(|\Omega(0)\rangle, \psi(\vec{x},0))$. In the Heisenberg picture this initial state evolves into $(|\Omega(0)\rangle, \psi(\vec{x},t))$. In the Schrödinger picture the state evolves into $(|\Omega(t)\rangle, \psi(\vec{x},0))$.

### 3. Gauge invariance in the Heisenberg picture

Now consider a "simple" field theory consisting of non-interacting electrons in the presence of a classical electromagnetic field. In this case the time evolution of the field operator in the Heisenberg picture is given by,

$$i\frac{\partial \hat{\psi}(\vec{x},t)}{\partial t} = H_D \hat{\psi}(\vec{x},t) \quad (3.1)$$

where,

$$H_D = H_0 - e\vec{\alpha}\cdot\vec{A} + eA_0 \quad (3.2)$$

and,

$$H_0 = -i\vec{\alpha}\cdot\vec{\nabla} + \beta m \quad (3.3)$$

In the above expression the electromagnetic potential $(A_0, \vec{A})$ is taken to be a classical, unquantized, real valued quantity. Also $e$ and $m$ are the charge and mass of the electron, respectively, and $\vec{\alpha}$ and $\beta$ are the usual 4x4 matrices. Note that in the above equations we use $\hbar = c = 1$.

Also assume that at the initial time $t = 0$ the Heisenberg field operator obeys the equal time anti-commutator relationships,

$$\hat{\psi}_\alpha^\dagger(\vec{x},0)\hat{\psi}_\beta(\vec{x}',0) + \hat{\psi}_\beta(\vec{x}',0)\hat{\psi}_\alpha^\dagger(\vec{x},0) = \delta_{\alpha\beta}\delta^{(3)}(\vec{x}-\vec{x}') \tag{3.4}$$

Define a Heisenberg operator by the expression,

$$\hat{O}_{op,H} = \hat{\psi}^\dagger(\vec{x},t) O_{op} \hat{\psi}(\vec{x},t) \tag{3.5}$$

The quantity $O_{op}$ operates on the field operator $\hat{\psi}(\vec{x},t)$. If $|\Omega(0)\rangle$ is a normalized state vector then the expectation value of the operator $\hat{O}_{op,H}$ in the Heisenberg picture is given by,

$$O_{op,H} = \langle\Omega(0)|\hat{O}_{op,H}|\Omega(0)\rangle = \langle\Omega(0)|\psi^\dagger(\vec{x},t) O_{op} \psi(\vec{x},t)|\Omega(0)\rangle \tag{3.6}$$

Next we will show that quantum field theory in the Heisenberg picture is gauge invariant. For a theory to be gauge invariant the expectation value of physical observables must be gauge independent. The physical observables that we will consider are the current and charge expectation values. The Heisenberg current and charge operators are, respectively defined by,

$$\hat{\vec{J}}_H(\vec{x},t) = e\hat{\psi}^\dagger(\vec{x},t)\vec{\alpha}\hat{\psi}(\vec{x},t) \text{ and } \hat{\rho}_H(\vec{x},t) = e\hat{\psi}^\dagger(\vec{x},t)\hat{\psi}(\vec{x},t) \tag{3.7}$$

The current and charge expectation values for a normalized state vector $|\Omega(0)\rangle$ are, defined by,

$$\vec{J}_{H,e}(\vec{x},t) = \langle\Omega(0)|\hat{\vec{J}}_H(\vec{x},t)|\Omega(0)\rangle; \quad \rho_{H,e}(\vec{x},t) = \langle\Omega(0)|\hat{\rho}_H(\vec{x},t)|\Omega(0)\rangle \tag{3.8}$$

It is easy to demonstrate gauge invariance in the Heisenberg picture. Assume that at the initial time $t = 0$ the initial state of the system is given by $(|\Omega(0)\rangle, \hat{\psi}(\vec{x}))$. Let the system evolve in time in the presence of an electromagnetic potential $(A_0, \vec{A})$. According to the Heisenberg picture the system evolves into $\left(|\Omega(0)\rangle, \hat{\psi}(\vec{x},t)\right)$ where $\hat{\psi}(\vec{x},t)$ satisfies,

$$i\frac{\partial \hat{\psi}(\vec{x},t)}{\partial t} = \left(H_0 - e\vec{\alpha}\cdot\vec{A} + eA_0\right)\hat{\psi}(\vec{x},t) \tag{3.9}$$

along with the initial condition $\hat{\psi}(\vec{x},0) = \hat{\psi}(\vec{x})$. Now suppose we start with the same system at the initial time $t=0$ and evolve in time in the presence of the gauge transformed potential $\left(A_{0,g}, \vec{A}_g\right) = \left(A_0 + \partial\chi/\partial t, \vec{A} - \nabla\chi\right)$ where $\chi(\vec{x},t)$ is an arbitrary real valued function that satisfies the initial condition,

$$\chi(\vec{x},0) = 0; \quad \left.\frac{\partial\chi(\vec{x},t)}{\partial t}\right|_{t=0} = 0 \tag{3.10}$$

In this case the initial system evolves into the system $\left(|\Omega(0)\rangle, \hat{\psi}_g(\vec{x},t)\right)$ where $\hat{\psi}_g(\vec{x},t)$ satisfies,

$$i\frac{\partial \hat{\psi}_g(\vec{x},t)}{\partial t} = \left(H_0 - e\vec{\alpha}\cdot\left(\vec{A} - \nabla\chi\right) + e\left(A_0 + \partial\chi/\partial t\right)\right)\hat{\psi}_g(\vec{x},t) \tag{3.11}$$

along with the initial condition $\hat{\psi}_g(\vec{x},0) = \hat{\psi}(\vec{x})$. It can easily be shown that,

$$\hat{\psi}_g(\vec{x},t) = e^{-ie\chi(\vec{x},t)}\hat{\psi}(\vec{x},t) \text{ and } \hat{\psi}_g^\dagger(\vec{x},t) = \hat{\psi}^\dagger(\vec{x},t)e^{ie\chi(\vec{x},t)} \tag{3.12}$$

It is evident that we obtain identical results when we substitute either $\hat{\psi}(\vec{x},t)$ or $\hat{\psi}_g(\vec{x},t)$ into (3.7). Then, referring to (3.8), it is evident that the current and charge expectation values in the Heisenberg picture do not depend on the gauge transformation and therefore the Heisenberg picture is gauge invariant.

## **4. The Schrödinger picture**

In the Schrödinger picture the state vector evolves in time according the Schrödinger equation,

$$i\frac{\partial |\Omega(t)\rangle}{\partial t} = \hat{H}|\Omega(t)\rangle \tag{4.1}$$

We can take the Hermitian conjugate of the above equation to obtain,

$$-i\frac{\partial \langle\Omega(t)|}{\partial t} = \langle\Omega(t)|\hat{H} \tag{4.2}$$

where,

$$\hat{H} = \int \hat{\psi}^\dagger(\vec{x},0) H_D \hat{\psi}(\vec{x},0) d\vec{x} \tag{4.3}$$

Next define a Schrödinger operator by the expression,

$$\hat{O}_{op,S} = \psi^\dagger(\vec{x},0) O_{op} \psi(\vec{x},0) \tag{4.4}$$

The expectation value of the Schrödinger operator $\hat{O}_{op,S}$ for the state vector $|\Omega(t)\rangle$ in the Schrödinger picture is given by,

$$O_{op,S}(t) = \langle \Omega(t)|\hat{O}_{op,S}|\Omega(t)\rangle \tag{4.5}$$

It can be shown that the expectation values in both pictures are the same, that is,

$$O_{op,S}(t) = O_{op,H}(t) \tag{4.6}$$

It is on this basis that the Heisenberg and Schrödinger pictures are considered to be equivalent representations of quantum theory. A proof that (4.6) is true is given in the Appendix.

## **5. The Vacuum state**

An expectation value is a number. Therefore in order to evaluate expectation values we need to know how the field operators act on the state vectors. We will start by assuming that at the initial time $t=0$ the state vector is in an initial unperturbed state which is given by,

$$\hat{\psi}_0(\vec{x}) = \sum_{\vec{p},s} \left( \hat{b}_{\vec{p},s} \phi_{1,\vec{p},s}(\vec{x}) + \hat{d}^\dagger_{\vec{p},s} \phi_{-1,\vec{p},s}(\vec{x}) \right); \; \hat{\psi}^\dagger_0(\vec{x}) = \sum_{\vec{p},s} \left( \hat{b}^\dagger_{\vec{p},s} \phi^\dagger_{1,\vec{p},s}(\vec{x}) + \hat{d}_{\vec{p},s} \phi^\dagger_{-1,\vec{p},s}(\vec{x}) \right) \tag{5.1}$$

where the $\hat{b}_{\vec{p},s}$ ($\hat{b}^\dagger_{\vec{p},s}$) are the destruction(creation) operators for an electron in the state $\phi_{1,\vec{p},s}(\vec{x})$ and $\hat{d}_{\vec{p},s}$ ($\hat{d}^\dagger_{\vec{p},s}$) are the destruction(creation) operators for an positron in the state $\phi_{-1,\vec{p},s}(\vec{x})$. They satisfy the anticommutator relation,

$$\hat{b}_{\vec{p},s}\hat{b}^\dagger_{\vec{p}',s'} + \hat{b}^\dagger_{\vec{p}',s'}\hat{b}_{\vec{p},s} = \delta_{s's}\delta^{(3)}(\vec{p}-\vec{p}'); \; \hat{d}_{\vec{p},s}\hat{d}^\dagger_{\vec{p}',s'} + \hat{d}^\dagger_{\vec{p}',s'}\hat{d}_{\vec{p},s} = \delta_{s's}\delta^{(3)}(\vec{p}-\vec{p}') \tag{5.2}$$

The $\phi_{\lambda,\vec{p},s}(\vec{x})$ are basis state solutions of the free field Dirac equation with energy eigenvalue $\lambda E_{\vec{p}}$ and can be expressed by

$$H_0 \phi_{\lambda,s,\vec{p}}(\vec{x}) = \lambda E_{\vec{p}} \phi_{\lambda,s,\vec{p}}(\vec{x}) \tag{5.3}$$

and where,

$$E_{\vec{p}} = +\sqrt{\vec{p}^2 + m^2}, \quad \lambda = \begin{cases} +1 \text{ for a positive energy state} \\ -1 \text{ for a negative energy state} \end{cases} \tag{5.4}$$

where $\vec{p}$ is the momentum of the state and $s = \pm 1/2$ is the spin index.

The $\phi_{\lambda,s,\vec{p}}(\vec{x})$ can be expressed by,

$$\phi_{\lambda,s,\vec{p}}(\vec{x}) = u_{\lambda,s,\vec{p}} e^{i\vec{p}\cdot\vec{x}} \tag{5.5}$$

where $u_{\lambda,s,\vec{p}}$ is a constant 4-spinor which are given in Chapt. 2 of Ref. [2]. The $\phi_{\lambda,s,\vec{p}}(\vec{x})$ form a complete orthonormal basis in Hilbert space and satisfy

$$\int \phi_{\lambda,s,\vec{p}}^\dagger(\vec{x}) \phi_{\lambda',s',\vec{p}'}(\vec{x}) d\vec{x} = \delta_{\lambda,\lambda'} \delta_{s,s'} \delta_{\vec{p},\vec{p}'} \tag{5.6}$$

Now that we have specified the field operator at the initial time we must define the state vectors on which the field operators act. First define the vacuum state $|0\rangle$ as the state that is destroyed by all electron and positron destruction operators, i.e.,

$$\hat{b}_{s,\vec{p}}|0\rangle = \hat{d}_{s,\vec{p}}|0\rangle = 0 \tag{5.7}$$

It can be shown that the vacuum state is an eigenstate of the Schrödinger free field energy operator which is defined by,

$$\hat{H}_{0,S} = \int \hat{\psi}_0^\dagger(\vec{x}) H_0 \hat{\psi}_0(\vec{x}) d\vec{x} \tag{5.8}$$

Using (5.1), (5.6), and (5.3) we can write the Schrödinger picture free field energy operator as,

$$\hat{H}_{0,S} = \sum_{s,\vec{p}} E_{\vec{p}} \left( b_{s,\vec{p}}^\dagger b_{s,\vec{p}} - d_{s,\vec{p}} d_{s,\vec{p}}^\dagger \right) \tag{5.9}$$

Then, using (5.7) and (5.2) we obtain,

$$\hat{H}_{0,S}|0\rangle = \varepsilon(|0\rangle)|0\rangle \tag{5.10}$$

where the eigenvalue $\varepsilon(|0\rangle)$ is given by,

$$\varepsilon(|0\rangle) = -\sum_{s,\vec{p}} E_{\vec{p}} \tag{5.11}$$

This is obviously a divergent quantity. However that will not be a problem because we are actually concerned with differences in the energy and not the actual value.

Additional eigenstates $|n\rangle$ are formed by acting on the vacuum state $|0\rangle$ with the various combinations of the creation operators $b_{s,\vec{p}}^\dagger$ and $d_{s,\vec{p}}^\dagger$. The effect of doing this is to create states with positive energy with respect to the vacuum state. The set of eigenstates $|n\rangle$ (which includes the vacuum state $|0\rangle$) form an orthonormal basis that satisfies the following relationships,

$$\hat{H}_{0,S}|n\rangle = \varepsilon(|n\rangle)|n\rangle \text{ where } \varepsilon(|n\rangle) > \varepsilon(|0\rangle) \text{ for } |n\rangle \neq |0\rangle \tag{5.12}$$

and

$$\langle n|m\rangle = \delta_{mn} \tag{5.13}$$

Any arbitrary normalized state $|\Omega\rangle$ can be expanded in terms of these basis states,

$$|\Omega\rangle = \sum_n c_n |n\rangle \tag{5.14}$$

where the normalization condition is expressed by,

$$\sum_n |c_n|^2 = 1 \tag{5.15}$$

The free field energy expectation value of this state is,

$$\langle \Omega|\hat{H}_{0,S}|\Omega\rangle = \sum_n |c_n|^2 \varepsilon(|n\rangle) \tag{5.16}$$

Use (5.12) and (5.15) to obtain the relationship,

$$\langle \Omega|\hat{H}_{0,S}|\Omega\rangle \geq \langle 0|\hat{H}_{0,S}|0\rangle = \varepsilon(|0\rangle) \text{ for all } |\Omega\rangle \tag{5.17}$$

This can be also written as,

$$\langle \Omega|\hat{H}_{0,S}|\Omega\rangle - \langle 0|\hat{H}_{0,S}|0\rangle \geq 0 \text{ for all } |\Omega\rangle \tag{5.18}$$

The key result of this section is that in the Schrödinger picture there is a lower bound to the free field energy of an arbitrary normalized state vector $|\Omega\rangle$.

### 6. Gauge invariance in the Schrödinger picture.

In Section 3 it was shown that the Heisenberg picture was gauge invariant. Here we shall consider the problem of gauge invariance in the Schrödinger picture. It will be shown that the Schrödinger picture is not gauge invariant. This will be done by assuming that the theory is gauge invariant and then finding a contradiction. The following discussion is similar to that given in Ref. [10].

First define the Schrödinger current and charge operators by,

$$\hat{\vec{J}}_S(\vec{x}) = e\hat{\psi}_0^\dagger(\vec{x})\vec{\alpha}\hat{\psi}_0(\vec{x}) \text{ and } \hat{\rho}_S(\vec{x}) = e\hat{\psi}_0^\dagger(\vec{x})\hat{\psi}_0(\vec{x}) \tag{6.1}$$

Using this along with (5.8), (4.3), and (3.2) we can write the Schrödinger Hamiltonian operator as,

$$\hat{H}(t) = \hat{H}_{0,S} - \int \hat{\vec{J}}_S(\vec{x}) \cdot \vec{A}(\vec{x},t) d\vec{x} + \int \hat{\rho}_S(\vec{x}) A_0(\vec{x},t) d\vec{x} \qquad (6.2)$$

Now at the initial time $t=0$ let the quantum state be given by the pair $(|\Omega_0\rangle, \psi_0(\vec{x}))$ where $\psi_0(\vec{x})$ is defined in (5.1) and $|\Omega_0\rangle$ with be specified shortly. Let the state evolve forward in time in the presence of an electromagnetic potential given by,

$$\left(A_0^{(1)}, \vec{A}^{(1)}\right) = 0 \qquad (6.3)$$

In the Schrödinger picture the system evolves into $(|\Omega_1(t)\rangle, \psi_0(\vec{x}))$ where $|\Omega_1(t)\rangle$ satisfies,

$$i\frac{\partial |\Omega_1(t)\rangle}{\partial t} = \hat{H}_{0,S} |\Omega_1(t)\rangle \qquad (6.4)$$

with the boundary condition $|\Omega_1(0)\rangle = |\Omega_0\rangle$. The solution to this equation is,

$$|\Omega_1(t)\rangle = e^{-i\hat{H}_{0,S} t} |\Omega(0)\rangle \qquad (6.5)$$

The current and charge expectation values are,

$$\vec{J}_{1,e}(\vec{x},t) = \langle \Omega_1(t)| \hat{\vec{J}}_S(\vec{x}) |\Omega_1(t)\rangle \qquad (6.6)$$

and,

$$\rho_{1,e}(\vec{x},t) = \langle \Omega_1(t)| \hat{\rho}_S(\vec{x}) |\Omega_1(t)\rangle \qquad (6.7)$$

Assume that we have chosen the initial state vector $|\Omega_0\rangle$ such that at some time $t_f > 0$ the quantity $\partial \rho_{1,e}(\vec{x}, t_f)/\partial t_f$ is nonzero. We can ensure this by specifying $|\Omega_0\rangle$ by,

$$|\Omega_0\rangle = \frac{1}{\sqrt{2}} \left( b^\dagger_{\vec{p}_1, s_1} + b^\dagger_{\vec{p}_2, s_2} \right) |0\rangle \qquad (6.8)$$

Use this in (6.5) to obtain,

$$|\Omega(t)\rangle = \frac{1}{\sqrt{2}} \left( b^\dagger_{\vec{p}_1, s_1} e^{-iE_{\vec{p}_1} t} + b^\dagger_{\vec{p}_2, s_2} e^{-iE_{\vec{p}_2} t} \right) |0\rangle \qquad (6.9)$$

Next use the above along with (5.1), and (6.7) to obtain,

$$\frac{\partial \rho_{1,e}(\vec{x},t)}{\partial t} = \frac{e}{2} \frac{\partial}{\partial t} \left( u^\dagger_{1, \vec{p}_1, s_1} u_{2, \vec{p}_2, s_2} e^{i(\vec{p}_2 - \vec{p}_1)\cdot \vec{x}} e^{-i(E_{\vec{p}_2} - E_{\vec{p}_1})t} + c.c. \right) \qquad (6.10)$$

where *c.c.* means take the complex conjugate of the preceding term. This quantity is, in general, non-zero. We can also show that,

$$\frac{\partial \rho_{1,e}(\vec{x},t)}{\partial t} + \nabla \cdot \vec{J}_{1,e}(\vec{x},t) = 0 \tag{6.11}$$

This is just the continuity equation which states that local charge is conserved.

Next start with the same initial system $\left(|\Omega_0\rangle, \psi_0(\vec{x})\right)$ at $t=0$ and let the state evolve forward in time in the presence of the electromagnetic potential given by

$$\left(A_0^{(2)}, \vec{A}^{(2)}\right) = \left(\frac{\partial \chi}{\partial t}, -\vec{\nabla}\chi\right) \tag{6.12}$$

where $\chi(\vec{x},t)$ is an arbitrary real valued function that satisfies the following initial condition (3.10) at t=0.

In this case the quantum systems evolves into the pair $\left(|\Omega_2(t)\rangle, \psi_0(\vec{x})\right)$ where $|\Omega_2(t)\rangle$ satisfies the initial condition $|\Omega_2(0)\rangle = |\Omega_0\rangle$ and obeys the Schrödinger equation (4.1) where, using (6.12) in (6.2), we write the Schrödinger picture Hamiltonian operator as,

$$\hat{H}(t) = \hat{H}_{0,S} + \int \hat{\vec{J}}_S(\vec{x}) \cdot \nabla \chi(\vec{x},t) d\vec{x} + \int \hat{\rho}_S(\vec{x}) \frac{\partial \chi(\vec{x},t)}{\partial t} d\vec{x} \tag{6.13}$$

Next consider the quantity $\langle \Omega_2(t)|\hat{H}_{0,S}|\Omega_2(t)\rangle$. Using (6.13) we can obtain the expression,

$$\langle \Omega_2(t)|\hat{H}_{0,S}|\Omega_2(t)\rangle = \langle \Omega_2(t)|\left(\hat{H}(t) - \int \hat{\vec{J}}_S(\vec{x}) \cdot \nabla \chi(\vec{x},t) d\vec{x} - \int \hat{\rho}_S(\vec{x}) \frac{\partial \chi(\vec{x},t)}{\partial t} d\vec{x}\right)|\Omega_2(t)\rangle$$

$$\tag{6.14}$$

Take the time derivative of the above equation and use,

$$\frac{\partial}{\partial t}\hat{H}(t) = \int \hat{\vec{J}}_S(\vec{x}) \cdot \nabla \frac{\partial \chi(\vec{x},t)}{\partial t} d\vec{x} + \int \hat{\rho}_S(\vec{x}) \frac{\partial^2 \chi(\vec{x},t)}{\partial t^2} d\vec{x} \tag{6.15}$$

to obtain,

$$\frac{\partial \langle \Omega_2(t)|\hat{H}_{0,S}|\Omega_2(t)\rangle}{\partial t} = -\int \frac{\partial \vec{J}_{2,e}(\vec{x},t)}{\partial t} \cdot \nabla \chi(\vec{x},t) d\vec{x} - \int \frac{\partial \hat{\rho}_{2,e}(\vec{x},t)}{\partial t} \frac{\partial \chi(\vec{x},t)}{\partial t} d\vec{x} \tag{6.16}$$

where $\vec{J}_{2,e}(\vec{x},t)$ and $\rho_{2,e}(\vec{x},t)$ are the current and charge expectation values, respectively, for the system $\left(\left|\Omega_2(t)\right\rangle,\psi_0(\vec{x})\right)$ and are given by,

$$\vec{J}_{2,e}(\vec{x},t)=\left\langle\Omega_2(t)\left|\hat{\vec{J}}_S(\vec{x})\right|\Omega_2(t)\right\rangle;\ \rho_{2,e}(\vec{x},t)=\left\langle\Omega_2(t)\left|\hat{\rho}_S(\vec{x})\right|\Omega_2(t)\right\rangle \qquad (6.17)$$

Now we will invoke the principle of gauge invariance. Note that the potentials $\left(A_0^{(1)},\vec{A}^{(1)}\right)$ and $\left(A_0^{(2)},\vec{A}^{(2)}\right)$ are related by a gauge transformation. If the theory is gauge invariant then the current and charge expectation values must be gauge invariant. Therefore,

$$\vec{J}_{2,e}(\vec{x},t)=\vec{J}_{1,e}(\vec{x},t) \text{ and } \rho_{2,e}(\vec{x},t)=\rho_{1,e}(\vec{x},t) \qquad (6.18)$$

Use this in (6.16) to obtain,

$$\frac{\partial\left\langle\Omega_2(t)\left|\hat{H}_{0,S}\right|\Omega_2(t)\right\rangle}{\partial t}=-\int\frac{\partial\vec{J}_{1,e}(\vec{x},t)}{\partial t}\cdot\nabla\chi(\vec{x},t)d\vec{x}-\int\frac{\partial\hat{\rho}_{1,e}(\vec{x},t)}{\partial t}\frac{\partial\chi(\vec{x},t)}{\partial t}d\vec{x} \qquad (6.19)$$

Next integrate the above equation by parts and rearrange terms to obtain,

$$\frac{\partial\left\langle\Omega_2(t)\left|\hat{H}_{0,S}\right|\Omega_2(t)\right\rangle}{\partial t}=\int\chi(\vec{x},t)\left(\frac{\partial\hat{\rho}_{1,e}(\vec{x},t)}{\partial t}+\nabla\cdot\vec{J}_{1,e}(\vec{x},t)\right)d\vec{x}-\frac{\partial}{\partial t}\int\frac{\partial\hat{\rho}_{1,e}(\vec{x},t)}{\partial t}\chi(\vec{x},t)d\vec{x}$$

(6.20)

Integrate this equation with respect to time from $t=0$ to $t=t_f$ and use (6.11) and the initial conditions (3.10) and $\left|\Omega_2(0)\right\rangle=\left|\Omega_0\right\rangle$ to obtain,

$$\left\langle\Omega_2(t)\left|\hat{H}_{0,S}\right|\Omega_2(t)\right\rangle-\left\langle\Omega_0\left|\hat{H}_{0,S}\right|\Omega_0\right\rangle=-\int\frac{\partial\hat{\rho}_{1,e}(\vec{x},t_f)}{\partial t_f}\chi(\vec{x},t_f)d\vec{x} \qquad (6.21)$$

Rearrange terms and subtract $\left\langle 0\left|\hat{H}_{0,S}\right|0\right\rangle$ from both sides to obtain,

$$\left\langle\Omega_2(t)\left|\hat{H}_{0,S}\right|\Omega_2(t)\right\rangle-\left\langle 0\left|\hat{H}_{0,S}\right|0\right\rangle=\Delta\xi-\int\frac{\partial\hat{\rho}_{1,e}(\vec{x},t_f)}{\partial t_f}\chi(\vec{x},t_f)d\vec{x} \qquad (6.22)$$

where,

$$\Delta\xi=\left\langle\Omega_0\left|\hat{H}_{0,S}\right|\Omega_0\right\rangle-\left\langle 0\left|\hat{H}_{0,S}\right|0\right\rangle \qquad (6.23)$$

Now in the above equation the quantities $\Delta\xi$ and $\hat{\rho}_{1,e}(\vec{x},t)$ are independent of $\chi(\vec{x},t)$. Therefore we can vary $\chi(\vec{x},t)$ in an arbitrary manner without affecting these other quantities. We will use this fact, along with the fact that $\partial \hat{\rho}_{1,e}(\vec{x},t_f)/\partial t_f$ is non-zero (see Eq. (6.10), to show that there is no lower bound to the quantity $\langle \Omega_2(t)|\hat{H}_{0,S}|\Omega_2(t)\rangle - \langle 0|\hat{H}_{0,S}|0\rangle$.

For example let $\chi(\vec{x},t_f) = f\left(\partial \hat{\rho}_{1,e}(\vec{x},t_f)/\partial t_f\right)$ where $f$ is an arbitrary constant. Use this in (6.21) to obtain,

$$\langle \Omega_2(t)|\hat{H}_{0,S}|\Omega_2(t)\rangle - \langle 0|\hat{H}_{0,S}|0\rangle = \Delta\xi - f\int\left(\frac{\partial \hat{\rho}_{1,e}(\vec{x},t_f)}{\partial t_f}\right)^2 d\vec{x} \qquad (6.24)$$

Now it should be evident that as $f \to \infty$ then $\left(\langle \Omega_2(t)|\hat{H}_{0,S}|\Omega_2(t)\rangle - \langle 0|\hat{H}_{0,S}|0\rangle\right) \to -\infty$. However this contradicts (5.18) which states that this quantity cannot be less than zero. Therefore there is an inconsistency between the requirement of gauge invariant and the relationship specified in (5.18). If (5.18) is true then the Schrödinger picture is not gauge invariant. This is consistent with the results of Ref. [9] where we proved the same result using a different approach. As discussed in [9] this inconstancy leads to the existence of non-gauge invariant terms when the polarization tensor is calculated.

## 7. Free field energy in the Heisenberg picture.

As we have shown the Schrödinger picture cannot be gauge invariant due to the fact that in the Schrödinger picture there is a lower bound to the free field energy as specified by (5.17) and (5.18). In this section we will consider the free field energy in the Heisenberg picture. Due to the fact that we have shown formally that the expectation values are the same in both picture we would expect that there is a lower bound to the free field energy in the Heisenberg picture. However we will show that, in contrast to the Schrödinger picture, there is no lower bound to the free field energy in the Heisenberg picture.

The Schrödinger free field energy operator was given in Eq. (5.8). The Heisenberg free field energy operator is given by,

$$\hat{H}_{0,H}(t) = \int \hat{\psi}^\dagger(\vec{x},t) H_0 \hat{\psi}(\vec{x},t) d\vec{x} \qquad (7.1)$$

Now at the initial time $t=0$ let the quantum state be given by the pair $\left(|\Omega_0\rangle, \psi_0(\vec{x})\right)$ where $\psi_0(\vec{x})$ has been defined by (5.1) and $|\Omega_0\rangle$ is given by (6.8). Let the system evolve forward in time with the electromagnetic potential given by (6.12) and (3.10).

In the Heisenberg picture the state evolves into $\left(|\Omega_0\rangle, \hat{\psi}(\vec{x},t)\right)$ where $\hat{\psi}(\vec{x},t)$ satisfies Eq. (3.1) and the initial condition $\hat{\psi}(\vec{x},0) = \hat{\psi}_0(\vec{x})$. Using (6.12) in (3.1) we obtain,

$$i\frac{\partial \hat{\psi}(\vec{x},t)}{\partial t} = \left(H_0 + e\vec{\alpha}\cdot\vec{\nabla}\chi + e\frac{\partial \chi}{\partial t}\right)\hat{\psi}(\vec{x},t) \tag{7.2}$$

The solution to the above equation is,

$$\hat{\psi}(\vec{x},t) = e^{-ie\chi}\hat{\psi}_0(\vec{x},t) \tag{7.3}$$

where,

$$\hat{\psi}_0(\vec{x},t) = e^{-iH_0 t}\hat{\psi}_0(\vec{x}) \tag{7.4}$$

In the Heisenberg picture the expectation value of the free field energy operator is $\langle \Omega_0 | \hat{H}_{0,H}(t) | \Omega_0 \rangle$. To evaluate this use (7.3) in (7.1) to obtain,

$$\hat{H}_{0,H}(t) = \int \hat{\psi}_0^\dagger(\vec{x},t) e^{+ie\chi} H_0 e^{-ie\chi} \hat{\psi}_0(\vec{x},t) d\vec{x} \tag{7.5}$$

Next use the following result,

$$H_0 e^{-ie\chi} \hat{\psi}_0(\vec{x},t) = e^{-ie\chi}\left(-e\vec{\alpha}\cdot\vec{\nabla}\chi + H_0\right)\hat{\psi}_0(\vec{x},t) \tag{7.6}$$

in (7.5) to obtain,

$$\hat{H}_{0,H}(t) = \int \hat{\psi}_0^\dagger(\vec{x},t)\left(-e\vec{\alpha}\cdot\vec{\nabla}\chi + H_0\right)\hat{\psi}_0(\vec{x},t) d\vec{x} \tag{7.7}$$

This yields,

$$\hat{H}_{0,H}(t) = \int \hat{\psi}_0^\dagger(\vec{x},t) H_0 \hat{\psi}_0(\vec{x},t) d\vec{x} - \int \hat{\vec{J}}_0(\vec{x},t)\cdot\vec{\nabla}\chi d\vec{x} \tag{7.8}$$

where,

$$\hat{\vec{J}}_0(\vec{x},t) = e\hat{\psi}_0^\dagger(\vec{x},t)\vec{\alpha}\hat{\psi}_0(\vec{x},t) \tag{7.9}$$

Use (7.4) to obtain,

$$\hat{H}_{0,H}(t) = \int \hat{\psi}_0^\dagger(\vec{x}) H_0 \hat{\psi}_0(\vec{x}) d\vec{x} - \int \hat{\vec{J}}_0(\vec{x},t)\cdot\vec{\nabla}\chi d\vec{x} = \hat{H}_{0,H}(0) - \int \hat{\vec{J}}_0(\vec{x},t)\cdot\vec{\nabla}\chi d\vec{x} \tag{7.10}$$

Sandwich the above between $\langle \Omega_0 |$ and $|\Omega_0\rangle$ to obtain,

$$\langle \Omega_0 | \hat{H}_{0,H}(t) | \Omega_0 \rangle = \langle \Omega_0 | \hat{H}_{0,H}(0) | \Omega_0 \rangle - \int \vec{J}_{0,e}(\vec{x},t)\cdot\vec{\nabla}\chi d\vec{x} \tag{7.11}$$

where,

$$\vec{J}_{0,e}(\vec{x},t) = \langle \Omega_0 | \hat{\psi}_0^\dagger(\vec{x},t)\vec{\alpha}\hat{\psi}_0(\vec{x},t) | \Omega_0 \rangle \tag{7.12}$$

Next assume reasonable boundary conditions and integrate by parts to obtain,

$$\langle\Omega_0|\hat{H}_{0,H}(t)|\Omega_0\rangle = \langle\Omega_0|\hat{H}_{0,H}(0)|\Omega_0\rangle + \int \chi(\vec{x},t)\vec{\nabla}\cdot\vec{J}_{0,e}(\vec{x},t)d\vec{x} \qquad (7.13)$$

Using (7.12), (6.8), and (5.1) we can show that,

$$\vec{\nabla}\cdot\vec{J}_{0,e}(\vec{x},t) = \frac{e}{2}\nabla\cdot\left(u^{\dagger}_{1,\vec{p}_1,s_1}\vec{\alpha}u_{2,\vec{p}_2,s_2}e^{i(\vec{p}_2-\vec{p}_1)\cdot\vec{x}}e^{-i(E_{\vec{p}_2}-E_{\vec{p}_1})t} + c.c.\right) \qquad (7.14)$$

Note that this is, in general, non-zero. Now recall that at the initial time $t=0$ the Schrödinger and Heisenberg operators are equal so that $\hat{H}_{0,H}(0) = \hat{H}_{0,S}$. Use this and subtract $\langle 0|\hat{H}_{0,S}|0\rangle$ from both sides of the above equation to obtain,

$$\langle\Omega_0|\hat{H}_{0,H}(t)|\Omega_0\rangle - \langle 0|\hat{H}_{0,S}|0\rangle = \Delta\xi + \int \chi(\vec{x},t)\vec{\nabla}\cdot\vec{J}_{0,e}(\vec{x},t)d\vec{x} \qquad (7.15)$$

where $\Delta\xi$ was defined in (6.23). Now based on (7.15) what can we say about $\langle\Omega_0|\hat{H}_{0,H}(t)|\Omega_0\rangle - \langle 0|\hat{H}_{0,S}|0\rangle$? The quantities $\vec{\nabla}\cdot\vec{J}_{0,e}(\vec{x},t)$ and $\Delta\xi$ are independent of $\chi(\vec{x},t)$. Therefore we can vary $\chi(\vec{x},t)$ without affecting these other quantities. For example, suppose we let $\chi(\vec{x},t) = -f\vec{\nabla}\cdot\vec{J}_{0,e}(\vec{x},t)$ where $f$ is a constant. In this case,

$$\langle\Omega_0|\hat{H}_{0,H}(t)|\Omega_0\rangle - \langle 0|\hat{H}_{0,S}|0\rangle = \Delta\xi - f\int\left|\vec{\nabla}\cdot\vec{J}_{0,e}(\vec{x},t)\right|^2 d\vec{x} \qquad (7.16)$$

As $f \to \infty$ we have $\left(\langle\Omega_0|\hat{H}_{0,H}(t)|\Omega_0\rangle - \langle 0|\hat{H}_{0,S}|0\rangle\right) \to -\infty$. Therefore there is no lower bound to the free field energy $\langle\Omega_0|\hat{H}_{0,H}(t)|\Omega_0\rangle$ in the Heisenberg picture.

If the expectation values in the Heisenberg picture are equal to those of the Schrödinger picture then we should be able to replace $\langle\Omega_0|\hat{H}_{0,H}(t)|\Omega_0\rangle$ in (5.18) with $\langle\Omega(t)|\hat{H}_{0,S}|\Omega(t)\rangle$ to obtain,

$$\langle\Omega_0|\hat{H}_{0,H}(t)|\Omega_0\rangle - \langle 0|\hat{H}_{0,S}|0\rangle = \langle\Omega(t)|\hat{H}_{0,S}|\Omega(t)\rangle - \langle 0|\hat{H}_{0,S}|0\rangle \qquad (7.17)$$

where $|\Omega(t)\rangle$ is the solution to the Schrödinger equation subject to the initial condition $|\Omega(0)\rangle = |\Omega_0\rangle$. Now according to (5.18) the right hand side of (7.17) must always be non-negative. However we have already shown that with proper selection of $\chi(\vec{x},t)$ and $|\Omega_0\rangle$ the quantity on the left hand side can be negative. Therefore we have a contradiction and the two pictures cannot be equivalent.

## 8. Summary of results.

A key result of this paper is that it shows that the Heisenberg and Schrödinger pictures are not equivalent. This is consistent with the work of Dirac [11][12] and runs counter to the widely held perception that two pictures are equivalent. In addition, we have tried to understand why non-gauge invariant terms appear in various calculations in QFT since the theory is supposed to be gauge invariant. The approach taken was to consider what is required for a "simple" field theory to be gauge invariance. It has been shown that in the Schrödinger picture the formal theory cannot be gauge invariant due to the fact that the vacuum state is a lower bound to the free field energy. Therefore calculations done using the Schrödinger picture as a starting point will yield non-gauge invariant results as is indeed the case. However if the theory is formulated in the Heisenberg picture it is easily shown to be gauge invariant at the formal level. This suggests that the problems of gauge invariance could be resolved by working in the Heisenberg picture instead of the Schrödinger picture.

Another important result is that there must be some kind of mathematical inconsistency in the theory. It was shown in the Appendix that the two pictures are formally equivalent. However on further examination we have shown that there is a lower bound to the free field energy in the Schrödinger picture but that there is no lower bound to the free field energy in the Heisenberg picture. We have also shown that the Heisenberg picture is gauge invariant while the Schrödinger picture is not. What accounts for differences between the two pictures which we have formally proved to be equivalent?

Now quantum theory is based on mathematics. A mathematical theory consists of postulates which are mathematical statements that are assumed to be true without proof. The postulates can then be used to prove additional mathematical statements called theorems. Now what does it imply if the theorems are not consistent with each other? It implies that the underlying postulates are not consistent.

The inconsistency in quantum field theory is due to the way the vacuum state is defined per the discussion in Section 5. The vacuum state is defined in such a way that in the Schrödinger picture it is a state of minimum free field energy as specified by Eqs. (5.17) and (5.18). However as was shown in Section 6 in order for the Schrödinger picture to be gauge invariant there must be no lower bound to the free field energy. Therefore there is a mathematical inconsistency in the Schrödinger picture between the requirement of gauge invariance and the requirement that the free field energy has a lower bound. As we have seen this also leads to the inequivalence between the Heisenberg and Schrödinger pictures even though these two pictures can be formally shown to be equivalent.

The next logical question to ask is whether it is possible to define the vacuum state in such a way that there is no lower bound to the free field energy in the Schrödinger picture? If this could be done then, perhaps, the problem of gauge invariance in the Schrödinger picture would be resolved. This question was address in Ref. [10]. As was shown in [10] it is, indeed, possible to define the vacuum so that there is no lower bound to the free field energy in the Schrödinger picture and when this is done QFT in the Schrödinger picture will be gauge invariant. This again emphasis the fact the failure of gauge invariance is due to the way the vacuum state is defined. When the vacuum state is defined as in [10] the Schrödinger picture will be gauge invariant.

In conclusion quantum field theory is mathematically inconsistent. This inconsistency manifests itself when quantities such as the vacuum current or polarization tensor are calculated. When these quantities are calculated non-gauge invariant terms appear in the result. These terms must be removed to obtain a physically correct solution.

**Appendix**

We will show that the expectation values in both pictures are equivalent. First, show that $\hat{\psi}(\vec{x},t)$ can be given by,

$$\hat{\psi}(\vec{x},t) = \hat{U}(t)^\dagger \hat{\psi}(\vec{x},0) \hat{U}(t) \tag{A.1}$$

where $\hat{U}(t)$ is an operator that evolves in time according to,

$$i\frac{\partial \hat{U}(t)}{\partial t} = \hat{H}\hat{U}(t) \tag{A.2}$$

where,

$$\hat{H} = \int \hat{\psi}^\dagger(\vec{x},0) H_D \hat{\psi}(\vec{x},0) d\vec{x} \tag{A.3}$$

and where $\hat{U}(t)$ satisfies the initial condition,

$$\hat{U}(0) = 1 \qquad (A.4)$$

Using the above, we can prove that $\hat{U}(t)$ is a unitary, that is, it satisfies,

$$\hat{U}(t)^\dagger = \hat{U}(t)^{-1} \qquad (A.5)$$

To prove this use that fact that $\hat{H}$ is hermitian along with (A.2) to obtain,

$$i\frac{\partial \hat{U}(t)^\dagger}{\partial t} = -\hat{U}(t)^\dagger \hat{H} \qquad (A.6)$$

Using this result along with (A.3) we obtain,

$$i\frac{\partial}{\partial t}\left(\hat{U}(t)^\dagger \hat{U}(t)\right) = -\left(\hat{U}(t)^\dagger \hat{H}\right)\hat{U}(t) + \hat{U}(t)^\dagger \left(\hat{H}\hat{U}(t)\right) = 0 \qquad (A.7)$$

Therefore $\hat{U}(t)^\dagger \hat{U}(t)$ is constant in time. Using this result and (A.4) we obtain $\hat{U}(t)^\dagger \hat{U}(t) = \hat{U}(0)^\dagger \hat{U}(0) = 1$. Therefore (A.5) is true.

To show that (A.1) is valid substitute (A.1) in (3.1), along with (A.5), to obtain,

$$i\frac{\partial\left(\hat{U}(t)^{-1}\hat{\psi}(\vec{x},0)\hat{U}(t)\right)}{\partial t} = H_D\left(\hat{U}(t)^{-1}\hat{\psi}(\vec{x},0)\hat{U}(t)\right) \qquad (A.8)$$

This yields

$$i\left[-\hat{U}^{-1}\frac{\partial \hat{U}}{\partial t}\hat{U}^{-1}\hat{\psi}(\vec{x},0)\hat{U} + \hat{U}^{-1}\hat{\psi}(\vec{x},0)\frac{\partial \hat{U}}{\partial t}\right] = H_D\left(\hat{U}^{-1}\hat{\psi}(\vec{x},0)\hat{U}\right) \qquad (A.9)$$

Use (A.2) in the above to obtain,

$$\left[-\hat{U}^{-1}\hat{H}\hat{U}\hat{U}^{-1}\hat{\psi}(\vec{x},0)\hat{U} + \hat{U}^{-1}\hat{\psi}(\vec{x},0)\hat{H}\hat{U}\right] = H_D\left(\hat{U}^{-1}\hat{\psi}(\vec{x},0)\hat{U}\right) \qquad (A.10)$$

Multiply the above equation by $\hat{U}$ from the left and $\hat{U}^{-1}$ from the right and use the fact that $\hat{U}$ commutes with $H_D$ and $\hat{U}\hat{U}^{-1} = 1$ to obtain,

$$\left[\hat{H}, \hat{\psi}(\vec{x},0)\right] = -H_D\hat{\psi}(\vec{x},0) \qquad (A.11)$$

We can use (3.4) to show that this equation is true. From (3.4) we obtain,

$$\left[\hat{\psi}^\dagger(\vec{x}',0) H_D \hat{\psi}(\vec{x}',0), \hat{\psi}(\vec{x},0)\right] = -\delta^{(3)}(\vec{x}' - \vec{x}) H_D \hat{\psi}(\vec{x}',0) \qquad (A.12)$$

Use this along with (A.3) to obtain,

$$[\hat{H},\hat{\psi}(\vec{x},0)] = -\int \delta^{(3)}(\vec{x}'-\vec{x})H_D\hat{\psi}(\vec{x}',0)d\vec{x}' = -H_D\hat{\psi}(\vec{x},0) \tag{A.13}$$

Therefore (A.11) is true which means that (A.1) and (A.2) are valid.

Next use (A.2) can be used to show that the solution to the Schrödinger equation (4.1) is given by,

$$|\Omega(t)\rangle = \hat{U}(t)|\Omega(0)\rangle \tag{A.14}$$

Use this result in (4.5) to show that expectation value of Schrödinger operator $\hat{O}_{op,S}$ for the state vector $|\Omega(t)\rangle$ in the Schrödinger picture is given by,

$$O_{op,S}(t) = \langle \Omega(0)|\hat{U}(t)^\dagger \hat{O}_{op,S}\hat{U}(t)|\Omega(0)\rangle \tag{A.15}$$

From (4.4) we obtain,

$$\hat{U}(t)^\dagger \hat{O}_{op,S}\hat{U}(t) = \hat{U}(t)^\dagger \psi^\dagger(\vec{x},0)O_{op}\psi(\vec{x},0)\hat{U}(t) \tag{A.16}$$

Next use $O_{op} = \hat{U}(t)\hat{U}(t)^\dagger O_{op} = \hat{U}(t)O_{op}\hat{U}(t)^\dagger$ along with (A.1) to obtain,

$$O_{op,S}(t) = \langle \Omega(0)|\psi^\dagger(\vec{x},t)O_{op}\psi(\vec{x},t)|\Omega(0)\rangle \tag{A.17}$$

Compare this result with (3.6) to obtain,

$$O_{op,H}(t) = O_{op,S}(t) \tag{A.18}$$

This shows that the Heisenberg and Schrödinger pictures are equivalent at the level of expectation values.